\documentclass[]{aa}
\usepackage{times,float}
\usepackage{epsfig}
\usepackage{graphics}
\usepackage{amssymb}
\usepackage{curves}
\usepackage{eepic}
\usepackage{epic}
\usepackage{aalongtable}
\usepackage{ulem}        

\begin{document}

\title{GRB\,021004 modelled by multiple energy injections
\thanks{Based on observations collected at CAHA, La Palma, 
Tirgo, USNO, Mt. John, Loiano, SAO and Plateau de Bure.}}
\titlerunning{GRB\,021004 modelled by multiple energy injections}
{\small\author{        
        A.~ de~ Ugarte~ Postigo \inst{1}   
   \and A.J.~ Castro-Tirado \inst{1}   
   \and J.~ Gorosabel \inst{1}
   \and G.~ J\'ohannesson \inst{2}
   \and G.~ Bj\"ornsson \inst{2}
   \and E.H.~ Gudmundsson \inst{2}
   \and M.~ Bremer \inst{3}
   \and S.~ Pak \inst{4}
   \and N.~ Tanvir \inst{4}
   \and J.M.~ Castro Cer\'on\inst{5}
   \and S.~ Guzyi \inst{6,1} 
   \and M.~ Jel\'{\i}nek \inst{1}
   \and S.~ Klose \inst{7}
   \and D.~ P\'erez-Ram\' irez \inst{8}
   \and J.~ Aceituno \inst{9}
   \and A.~ Campo Bagat\'{\i}n \inst{10}
   \and S.~ Covino \inst{11}
   \and N.~ Cardiel \inst{9}
   \and T.~ Fathkullin \inst{12}
   \and A.A.~ Henden \inst{13}
   \and S.~ Huferath \inst{14}
   \and Y.~ Kurata \inst{15}
   \and D.~ Malesani \inst{11}
   \and F.~ Mannucci \inst{16}
   \and P.~ Ruiz-Lapuente \inst{17}
   \and V.~ Sokolov \inst{12}
   \and U.~ Thiele \inst{9}
   \and L.~ Wisotzki \inst{14}
   \and L.A.~ Antonelli \inst{18}
   \and C.~ Bartolini \inst{19}
   \and A.~ Boattini \inst{20}
   \and A.~ Guarnieri \inst{19}
   \and A.~ Piccioni \inst{19}
   \and G.~ Pizzichini \inst{21}
   \and M.~ del Principe \inst{22}
   \and A.~ di~ Paola \inst{18}
   \and D.~ Fugazza \inst{23}
   \and G.~ Ghisellini \inst{11}
   \and L.~ Hunt \inst{16}
   \and T.~ Konstantinova \inst{24}
   \and N.~ Masetti \inst{21}
   \and E.~ Palazzi \inst{21}
   \and E.~ Pian \inst{25}
   \and M.~ Stefanon \inst{11}
   \and V.~ Testa \inst{18}
   \and P.J.~ Tristram \inst{26}.}
}

\institute{ 
           Instituto de Astrof\'{\i}sica de Andaluc\'{\i}a (IAA-CSIC),
           Apartado de Correos 3004, E-18080 Granada, Spain.
           \and  
           Science Institute, University of Iceland, 
           Dunhaga 3, IS-107 Reykjav\'{\i}k, Iceland.{}
           \and
           IRAM - Institute de Radio Astronomie Millim\'etrique,
           300 Rue de la Piscine, F-38406 Saint-Martin d'H\`eres, France.{}
           \and
           Dept. Physics, Astronomy \& Maths, University of Hertfordshire,
           College Lane, Hatfield, Herts, AL10 9AB, United Kingdom.{}
           \and
           Niels Bohr Institute, University of Copenhagen, 
           Juliane Maries Vej 30, DK-2100 Copenhagen \O{}, Denmark.{}
           \and
           Nikolaev State University, Nikolska 24, 54030, Nikolaev, Ukraine.{}
           \and
           Th\"uringer Landessternwarte Tautenburg, D-07778 Tautenburg, Germany.{}
           \and
           Dpto. de F\'{\i}sica (EPS), Universidad de Ja\'en, E-23071, Ja\'en, Spain.{}
           \and
           Calar Alto Observatory, Apartado de Correos 511, E-04080, Almer\'{\i}a, Spain.{}
           \and
           Departamento de F\'{\i}sica, Universidad de Alicante, Apartado de Correos 99, E-03080 Alicante, Spain.{}
           \and
           INAF - Osservatorio Astronomico di Brera, Via E. Bianchi 46, I-23807 Merate (LC), Italy.{}
           \and
           Special Astrophysical Observatory of the Russian Academy of Sci. (SAO-RAS) , Russia.{}
           \and
           U.S. Naval Observatory, Flagstaff, AZ 86001, USA.{}
           \and
           Astrophysikalisches Institut, D-14482 Potsdam, Germany.{}
           \and
           Solar Terrestrial Environment Laboratory, Nagoya University, Japan.{}
           \and
           INAF - IRA, Largo E. Fermi 5, I-50125 Firenze, Italy.{}
           \and
           Departamento de Astronom\'{\i}a, Universidad de Barcelona, Mart\'{\i} i Franqu\`es 1, E-08028 Barcelona, Spain.{}
           \and
           INAF - Osservatorio Astronomico di Roma, Via Frascati 33, I-00040 Monteporzio Catone, Italy.{}
           \and
           Dipartimento di Astronomia, Universit\'a di Bologna, Via Ranzani 1, I-40127 Bologna, Italy.{}
           \and
           INAF - IAS, Via Fosso del Cavaliere 100, I-00133 Roma, Italy.{}
           \and
           INAF - Istituto di Astrofisica Spaziale e Fisica Cosmica, Sezione di Bologna, Via Gobetti 101, I-40129 Bologna, Italy.{}
           \and
           INAF - Osservatorio Astronomico di Teramo, Via M. Maggini 47, I-64100 Teramo, Italy.{}
           \and
           Centro Galileo Galilei, Apartado 565, E-38700 Santa Cruz de La Palma, Spain.{}
           \and
           Astronomical Institute of St. Petersburg University, Petrodvorets, Universitetsky pr. 28, 198504 St. Petersburg, Russia.{}
           \and
           INAF - Osservatorio Astronomico di Trieste, Via Tiepolo 11, I-34131 Trieste, Italy.{}
           \and
           Department of Physics and Astronomy, University of Canterbury, Canterbury, New Zealand.{}
}

\offprints{ \hbox{A. de Ugarte Postigo, ({\tt  
deugarte@iaa.es})}}

\date{Received  / Accepted }

\abstract{
   
   GRB\,021004 is  one of  the best  sampled gamma-ray bursts (GRB)  to date, 
although the nature of its light curve is still being debated. Here we present 
a large amount  (107) of  new optical, near-infrared (NIR) and  
millimetre observations, ranging from 2 hours to more than a year after the burst. 
Fitting the multiband data to a model based on multiple energy injections
suggests that at least 7 refreshed shocks took place during the 
evolution of the  afterglow, implying a total energy release (collimated within 
an angle of 1$\fdg$8) of $\sim$ 8 $\times$ 10$^{51}$ erg. Analysis of the 
late photometry reveals that the GRB\,021004 host is a low extinction 
($A_{V}   \sim   0.1$) starburst galaxy with $M_{B}\simeq-22.0$.
 
 }\maketitle

\section{Introduction}

\vspace{0.3cm}
   
  At 12:06:13.57 UT  4th October 2002 a long-duration GRB 
triggered the instruments aboard the HETE-2 satellite. The detection
was  immediately transmitted  to observatories  all around  the  globe that
began observing a few minutes after the burst. A fast identification of the
optical afterglow  (Fox \cite{Fox02}) allowed observations of the event  
from the  first stages,  producing  one of  the best  multiwavelength coverage 
of a GRB obtained to date.

   Here we present a compilation of observations covering visible, NIR
and  millimetre wavelengths. We revisit the light curve of GRB\,021004 
using new data together with previously published data. Our  study covers
almost the  complete history  of the  event, from a  few minutes  after the
trigger to more than a year after, when the afterglow light disappeared 
into the underlying galaxy. We pay special  attention to the bumpy nature 
of the light curve and, using the best multiwavelength sampling to date, 
apply the multiple   energy    injection   model   described    by   
Bj\"ornsson et al. (\cite{Bjor04}).

   In Sect. 2 we present the observations and the methods 
that were used for the data reduction. Sect. 3 gives an introduction to the 
model we used for fitting the data. Sect. 4 presents a study 
of the extinction derived from the spectral flux distribution, the 
modeling of the afterglow and the properties of the host galaxy. 
Sect. 5 discusses the implications of the modeling proposed here. 
In Sect. 6 we present our conclusions.

Throughout, we assume a cosmology where $\Omega_{\Lambda}=0.7$,
$\Omega_{M}=0.3$ and $H_0=72$ km s$^{-1}$ Mpc$^{-1}$. Under these 
assumptions, the luminosity distance of GRB\,021004 is $d_l=18.2$ Gpc and
the look-back time is 10.4~Gyr (79.5 \% of the present Universe age).

\section{Observations and data reduction}

\subsection{Optical and NIR observations}

For this data set we have made use  of 11 telescopes, 9 in optical bands 
and 2 in NIR bands.  The observations 
started  2 hours after the burst and  extended to more than a year later. 
The  images where reduced with standard  procedures based  on 
IRAF\footnote{IRAF  is  distributed   by  the  National  
Optical  Astronomy Observatories,  which is operated  by the  Association 
of  Universities for Research in  Astronomy, Inc.  (AURA)  under cooperative 
agreement  with the National   Science   Foundation.}    and   JIBARO (de 
Ugarte Postigo et al. \cite{Deug05}). 

\begin{figure}[ht!]
\centering
\includegraphics[width=\hsize]{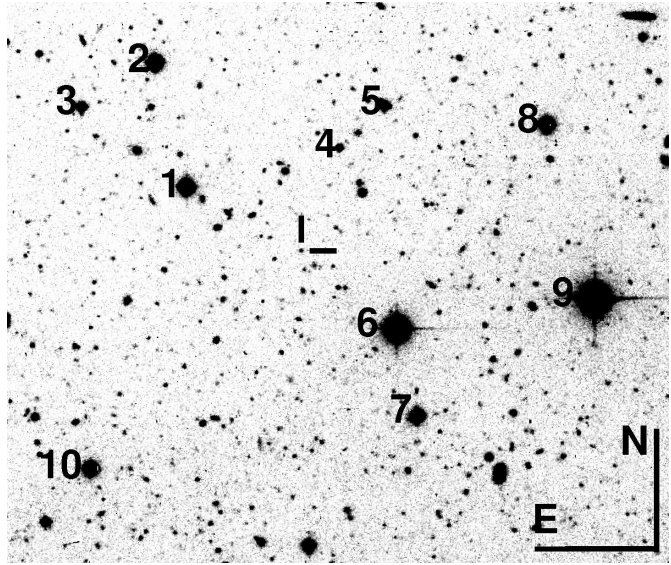}
\protect\caption[ ]{Reference stars in the surroundings of the GRB\,021004 
location (bracketed into tick marks). The depicted field of view is 
4.3$'$ x 4.0$'$. \label{calibfig}}
\end{figure}

Photometric  calibration of the optical images is based  on Henden 
(\cite{Hend02}), while the NIR calibration  was carried  out observing  
NIR standards
(Persson et al. \cite{Pers98}) at a  similar airmass as the GRB field.  The
instrumental magnitudes obtained were based on aperture photometry running
under  DAOPHOT.   Table~\ref{calibtab}  displays the  magnitudes  of  the
secondary standards shown in Fig.~\ref{calibfig}. The magnitude errors was
calculated by adding  in quadrature  the zero point  error (obtained  from the
dispersion of the secondary  standards) and the afterglow statistical error
given by DAOPHOT.  Table~ \ref{visnir} displays the complete optical/NIR list 
of observations performed by our collaboration on this event.

\begin{table}[h]
\begin{center}
\caption{Calibration stars in the GRB\,021004 field, marked in Fig. 1. \label{calibtab}} \scriptsize{
\begin{tabular}{r c c c c}
\noalign{\smallskip} \hline\hline \noalign{\smallskip}
&U&B&V&R\\
\noalign{\smallskip} \hline \noalign{\smallskip}
1&17.37$\pm$0.03&17.34$\pm$0.03&16.70$\pm$0.01&16.33$\pm$0.01\\
2&18.54$\pm$0.04&17.43$\pm$0.03&16.26$\pm$0.01&15.54$\pm$0.02\\
3&20.70$\pm$0.04&19.48$\pm$0.02&17.90$\pm$0.02&17.06$\pm$0.02\\
4&---&21.12$\pm$0.11&19.74$\pm$0.04&18.78$\pm$0.06\\
5&---&19.76$\pm$0.03&18.29$\pm$0.02&17.36$\pm$0.03\\
6&16.49$\pm$0.05&15.51$\pm$0.03&14.44$\pm$0.05&13.88$\pm$0.06\\
7&18.10$\pm$0.02&18.09$\pm$0.01&17.49$\pm$0.01&17.14$\pm$0.01\\
8&17.83$\pm$0.16&17.61$\pm$0.02&16.71$\pm$0.01&16.20$\pm$0.02\\
9&14.71$\pm$0.09&14.62$\pm$0.08&13.97$\pm$0.05&13.62$\pm$0.08\\
10&17.62$\pm$0.06&17.84$\pm$0.03&17.32$\pm$0.01&17.00$\pm$0.01\\
\noalign{\smallskip} \hline\hline \noalign{\smallskip}
&I&J&H&K\\
\noalign{\smallskip} \hline \noalign{\smallskip}
1&15.95$\pm$0.02&15.48$\pm$0.07&15.16$\pm$0.10&14.90$\pm$0.14\\
2&14.90$\pm$0.03&14.06$\pm$0.03&13.47$\pm$0.03&13.36$\pm$0.05\\
3&16.08$\pm$0.05&15.01$\pm$0.04&14.46$\pm$0.06&14.18$\pm$0.08\\
4&17.83$\pm$0.06&---&---&---\\
5&16.46$\pm$0.03&15.41$\pm$0.06&14.72$\pm$0.06&14.44$\pm$0.10\\
6&13.39$\pm$0.07&12.56$\pm$0.02&12.07$\pm$0.02&11.95$\pm$0.02\\
7&16.79$\pm$0.03&16.33$\pm$0.13&15.84$\pm$0.16&---\\
8&15.70$\pm$0.03&14.96$\pm$0.04&14.43$\pm$0.05&14.61$\pm$0.11\\
9&13.25$\pm$0.09&12.61$\pm$0.02&12.23$\pm$0.02&12.18$\pm$0.02\\
10&16.64$\pm$0.02&16.20$\pm$0.11&15.75$\pm$0.14&---\\
\noalign{
\smallskip} \hline \hline\end{tabular}
} \normalsize \rm
\end{center}
\end{table}

\subsection{Millimetre observations}

The dataset is completed with observations obtained in 230 GHz and 
90 GHz bands (see Table~ \ref{mmtab}) at the 6-antenna Plateau de Bure 
Interferometer
(PdB, Guilloteau et al. \cite{Guil92}). Data calibration was done with CLIC
and the UV plane fitting and analysis with MAPPING, which are part
of the GILDAS software package\footnote{GILDAS is the software package 
distributed by the IRAM Grenoble GILDAS group.}. MWC349 was used as primary 
flux calibrator and 0109+224 as phase calibrator.

\begin{table}[h]
\begin{center}
\caption{Millimetre observations of the  GRB\,021004 afterglow 
at PdB. \label{mmtab} }\scriptsize{
\begin{tabular}{c c c c}
\noalign{\smallskip} \hline
\hline \noalign{\smallskip}
Date 2002  & Frequency  & Flux  & Flux Error \\
  (UT)     &   (GHz)    & (mJy) &   (mJy)    \\
\noalign{\smallskip} \hline \noalign{\smallskip}
Oct 5.9844 &   86.293   & 2.47  &    0.29    \\
Oct 5.9844 &  231.700   & 1.22  &    1.22    \\
Oct 6.1458 &  115.261   & 1.62  &    1.44    \\
Oct 6.1458 &  231.700   & 0.22  &    3.65    \\
Oct 7.1438 &   87.717   & 2.57  &    0.56    \\
Oct 7.1438 &  232.034   & 3.26  &    1.54    \\
Oct 10.981 &   86.235   & 1.67  &    0.34    \\
Oct 10.981 &  208.475   & 4.71  &    1.96    \\
Oct 19.919 &   92.016   & 0.97  &    0.25    \\
Oct 19.919 &  231.972   & 1.60  &    1.00    \\
Nov 5.9813 &   97.991   & 0.15  &    0.27    \\
Nov 5.9813 &  239.551   &-0.33  &    0.71    \\
\noalign{
\smallskip} \hline \hline\end{tabular}
} \normalsize \rm
\end{center}
\end{table}

\section{Brief description of the modelling}

Our starting point is the standard fireball model 
(see e.g.\ Piran \cite{Piran2005}). 
To account for the observed light curve brightenings, we
modify the model by adding multiple energy injection
episodes (see Bj\"ornsson et al. \cite{Bjor04} and in particular
J\'ohannesson et al. \cite{Joha05} for a detailed discussion 
of the expressions and formulae used). We assume that the
central engine releases, essentially simultaneously, several shells
with different Lorentz factors. The evolution of the fireball is then
derived, as in Rhoads (\cite{Rhoa99}), from the conservation of energy and
momentum. The fastest moving shell drives the initial evolution of the
afterglow, but as it decelerates, the slower moving shells catch up
with the shock front, producing an energy injection.  Each shell collision is
assumed to be instantaneous and the dynamics of the interaction is
neglected, as well as any reverse shock contribution (these are
expected to contribute mostly at early stages in the fireball
evolution).

As in the standard fireball model, the afterglow radiation is assumed
to be of synchrotron origin and the local spectrum at each point in
the radiating shell is approximated by smoothly joined power law
segments (similar to Granot \& Sari \cite{Gran02}). 
Assuming that the shell is homogeneous in the co-moving frame,
its thickness is obtained from the shock conditions (Blandford 
\& McKee \cite{Blan76}) and from the conservation of swept up 
particles. The total flux at a given frequency and observer time is
then obtained by integrating over the equal arrival time surface
(Granot et al. \cite{Gran99} and references therein). 
The polarization light curve and position angle can be calculated 
adapting the model of Ghisellini \& Lazzati (\cite{Ghis99}) to the 
fireball model (see Bj\"ornsson et al. \cite{Bjor04}).

\section{Results}

\label{results}

\subsection{Multiwavelength light curves}

\begin{figure}[ht!]
\begin{center}
  {\includegraphics[width=\hsize]{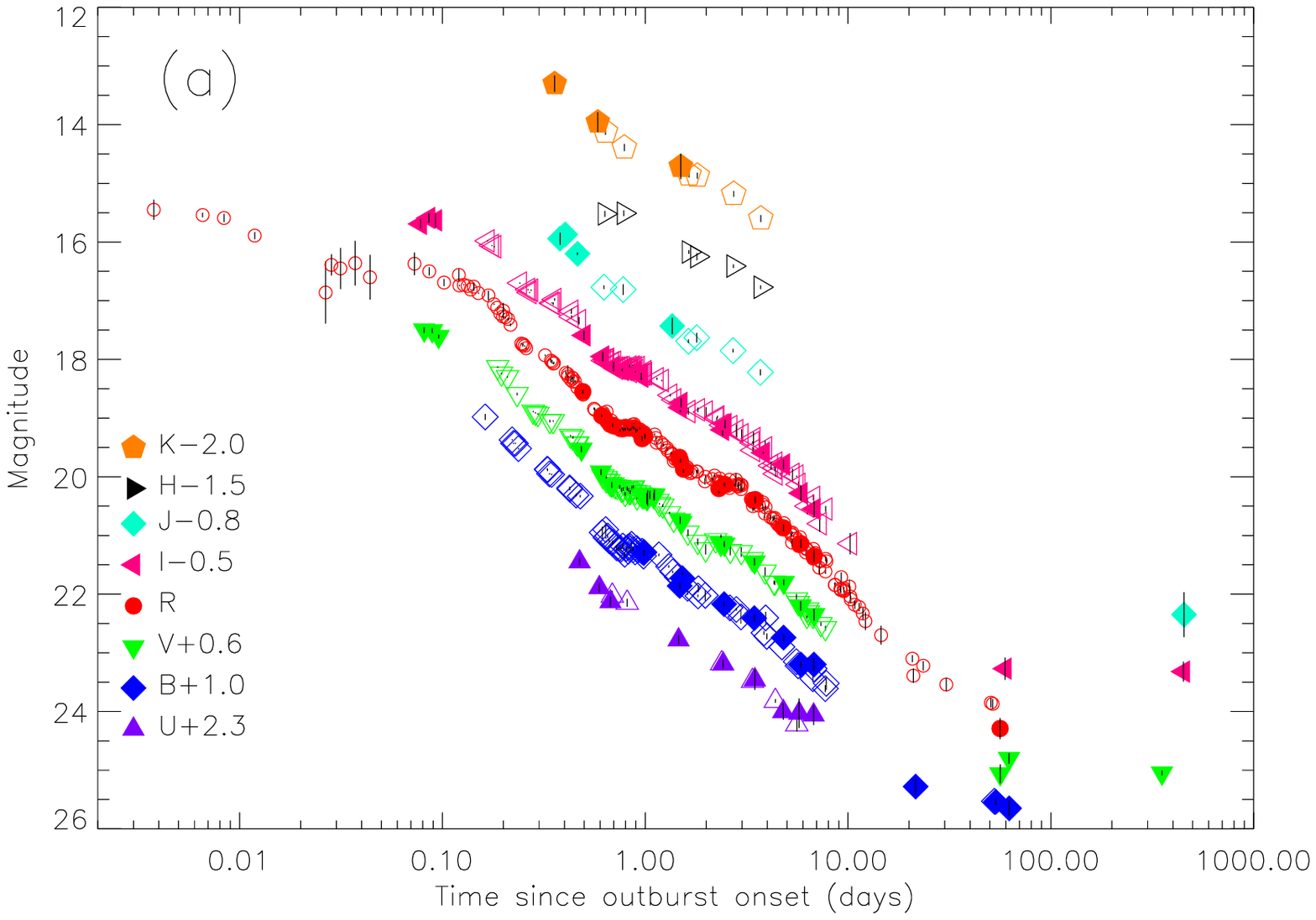}}
  {\includegraphics[width=\hsize]{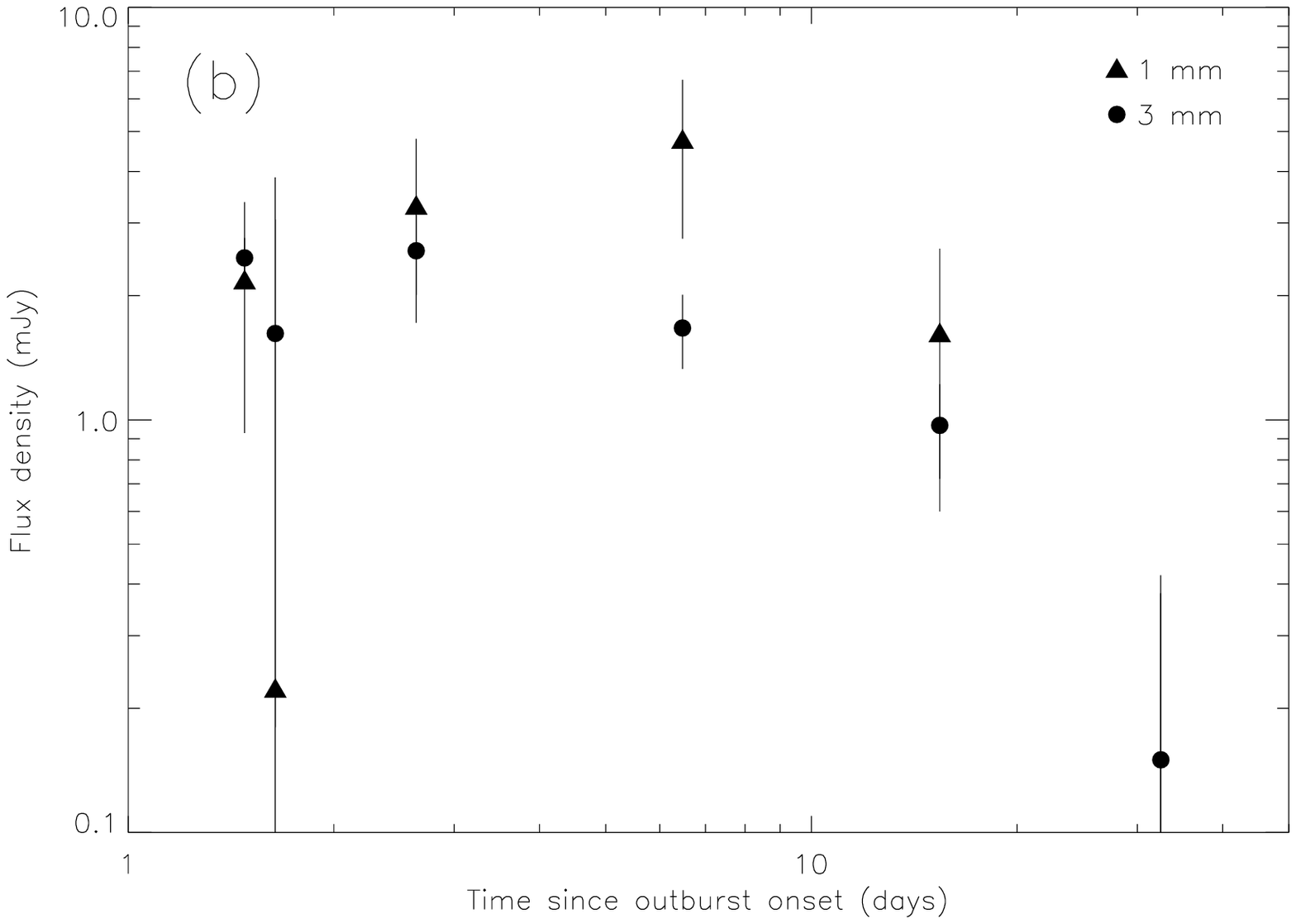}}
\caption{\label{lightcurve}  {\bf a)}  Optical  and  NIR  light  curves  
of  GRB\,021004 for the first $\sim$~500 days after the event. The different 
bands have been  intentionally separated for clarity.  Our observations  
are marked with filled  symbols while  published data are  represented with  
void ones (see text for references). {\bf b)} Millimetre light curves for 
the first $\sim$~35 days, obtained at PdB.}
\end{center}
\end{figure}

Fig.~\ref{lightcurve} shows  the light curves in the visible, NIR and
millimetre bands for GRB\,021004. The optical/NIR data points are plotted
together  with other  published  data (Fox  et  al.  \cite{Fox03};  Uemura 
et al. \cite{Uemu03}; Pandey et al. \cite{Pand03}; Bersier et al. \cite{Bers03}; 
Holland et al. \cite{Holl03}; Mirabal et al. \cite{Mira03}; Pak et al. 
\cite{Pak05}) in order to  show the  complexity of  the light curves.
 
\subsection{The optical and NIR SFD}
\label{Avconstrain}

As  a starting  point  all  the optical/NIR  magnitudes  are corrected  for
foreground Galactic extinction ($E{(B-V)}=0.06$; Schlegel  et al.  
\cite{Sche98}). Then, we
estimate the restframe extinction ($A_{V}$) and the favoured extinction law
based  on  the afterglow  optical/NIR  spectral flux distribution  (SFD)
constructed for several epochs.  The  selected epochs are those for which a
quasi-simultaneous wide optical/NIR coverage is available.  The SFDs are 
clustered around 9 epochs  displayed  in  Table~\ref{sedstable}. For each 
subset of photometric measurements we subtract the underlying host galaxy 
(see Sect.~\ref{host}).   This contribution is  significant only after the
first week.   Finally, we fit each SFD by using a  power law dimmed with  
different extinction laws (Pei \cite{Pei92}): Milky Way (MW),  Large 
Magellanic Cloud (LMC) and Small Magellanic Cloud (SMC).

We obtain  the best $\chi^{2}/d.o.f.$ (where d.o.f.\ stands for degree 
of freedom) with  the SMC extinction  law (see Table~\ref{chistable}), 
as it has been  previously observed for other GRB afterglows (Jensen et 
al. \cite{Jens01}; Fynbo et al. \cite{Fynb01}; Holland et al. \cite{Holl03}).
For  each   epoch  the  spectral   slope  ($\beta$;  the flux being   
$F_{\nu}  \propto\nu^{-\beta}$) and $A_{V}$ are calculated.  We can adopt 
the averaged SMC values for $A_{V}$ and $\beta$ since there is no evolution 
of the SFD on the considered time interval. The mean value inferred for  
the extinction and spectral index  are, $\langle A_{V}\rangle = 0.20 \pm 0.08$, 
and $\langle\beta\rangle = 0.5 \pm 0.2$, respectively.

We note that the unextinguished SFD  in the optical/NIR  range might not 
be well represented by a perfect power law spectrum, showing some degree 
of intrinsic convex curvature (see the shape of the spectra in Fig. 4). 
Thus,  the $A_{V}$ values displayed in  
Table~\ref{seds} have to be considered  a formal  upper limit,  likely 
close to  the real  ones. The inferred  $\langle A_{V}\rangle$  is used  
as  the  starting  point for  correcting  the intrinsic extinction of the 
object when applying the model.

\begin{table}[h]
\begin{center}
\caption{The GRB\,021004 SFD at 9 epochs, an SMC extinction law has been
assumed.\label{sedstable}} \scriptsize{
\begin{tabular}{c c c c c}
\noalign{\smallskip} \hline\hline \noalign{\smallskip}
SFD \# & Time since outburst & $\beta$ & $A_{V}$ & $\chi^{2}/d.o.f.$ \\
       & onset (days) & & & \\
\noalign{\smallskip} \hline \noalign{\smallskip}
  1   & 0.3609 & 0.43$\pm$0.18 & 0.17$\pm$0.04 & 0.8 \\
  2   & 0.6380 & 0.30$\pm$0.07 & 0.24$\pm$0.02 & 3.1 \\
  3   & 0.7851 & 0.20$\pm$0.10 & 0.29$\pm$0.04 & 1.5 \\
  4   & 1.4216 & 0.47$\pm$0.24 & 0.17$\pm$0.06 & 1.7 \\
  5   & 1.6304 & 0.82$\pm$0.14 & 0.08$\pm$0.06 & 0.3 \\
  6   & 1.8090 & 0.47$\pm$0.09 & 0.24$\pm$0.03 & 1.3 \\
  7   & 2.7018 & 0.39$\pm$0.08 & 0.26$\pm$0.03 & 0.6 \\
  8   & 3.6520 & 0.78$\pm$0.10 & 0.09$\pm$0.04 & 2.6 \\
  9   & 5.7388 & 0.47$\pm$0.20 & 0.25$\pm$0.06 & 0.4 \\
\noalign{
\smallskip} \hline \hline\end{tabular}
} \normalsize \rm
\end{center}
\end{table}

\begin{table}[h]
\begin{center}
\caption{Mean $\chi^{2}/d.o.f.$ obtained by fitting each extinction law on
the 9 available SFDs.\label{chistable}} \scriptsize{
\begin{tabular}{c c c c c}
\noalign{\smallskip} \hline\hline \noalign{\smallskip}
Extinction Law    &  NE  &  MW  &  LMC  &  SMC  \\
\noalign{\smallskip} \hline \noalign{\smallskip}
$\langle\chi^{2}/d.o.f.\rangle$ & 11$\pm$7 & 11$\pm$9 &  3.2$\pm$1.9  &  2.0$\pm$1.4  \\
\noalign{
\smallskip} \hline \hline\end{tabular}
} \normalsize \rm
\end{center}
\end{table}

\subsection{Afterglow model}

   A number of attempts have been made to explain  the nature of the  
bumps seen in this GRB's light curve (Lazzati et al. \cite{Lazz02}; 
Schaefer et al. \cite{Scha03}; Nakar et al. \cite{Naka03}).

In the present work we  show that the light curve can also
be described by multiple energy injections, using the model of Bj\"ornsson 
et al. (\cite{Bjor04}). Our multiwavelength data is fitted along with other 
measurements reported in the literature, optical and NIR data cited in 
Sect. 4.1 together with X-ray data from Sako \& Harrison (\cite{Sako02a}, 
\cite{Sako02b}) and radio data from Berger et al. (\cite{Berg02}) 
and Frail \& Berger (\cite{Frai02}). The model only  
reproduces the afterglow, hence the  contribution of the host galaxy  has 
been  subtracted (see Sect.~\ref{host}).

This GRB is located at a redshift of \textit{z}=2.3293 (Castro-Tirado et al. 
\cite{ajct05}) which shifts the Lyman-$\alpha$ break to the range of the 
\textit{U}-band. Thus, we must consider a correction for the Lyman-$\alpha$ 
blanketing that appears at shorter wavelengths. We use the model described by 
Madau (\cite{Mada95}) at this redshift and convolve it with the Johnson 
\textit{U}-band. This yields a reduction of the measured flux to 82\% of 
the original one. Due to the uncertainty of this approximation we do not use 
the corrected \textit{U}-band for fitting the model, but only for the 
verification of it.

From the analysis of the SFD done in the optical/NIR range 
(Sect.~\ref{Avconstrain}), an  SMC extinction  law with of  $A_{V} \lesssim
0.2$ is  favoured. The multiband  fitting has been  tested using a  grid of
extinctions ranging from zero to  $A_{V} = 0.4$ (within 2.5 sigma of the best
fit value obtained from the optical/NIR SFD fitting).  After several iterations
we find that the best fit of the whole multi-range data set is achieved with
$A_{V} \sim 0.1$.

The parameters that result from the best fit of our model  are displayed in 
Table~  \ref{tabmodel}.  The fitted  model is  characterized by  an initial  
shock followed  by 7 subsequent refreshed shocks, the last injection being 
the most energetic. The number of injection episodes is higher than in
Bj\"ornsson et al. (\cite{Bjor04}), as a result of a more complete dataset. 
Two injections are needed to account for late time radio data, and one 
($E_2$) is added to better model an optical bump at $\approx 0.35$ days. In 
addition, the electron energy index $p$ is a free parameter here, but was 
fixed in Bj\"ornsson et al. (\cite{Bjor04}).

\begin{figure*}[t]
\begin{center}
  {\includegraphics[height=22cm,angle=0]{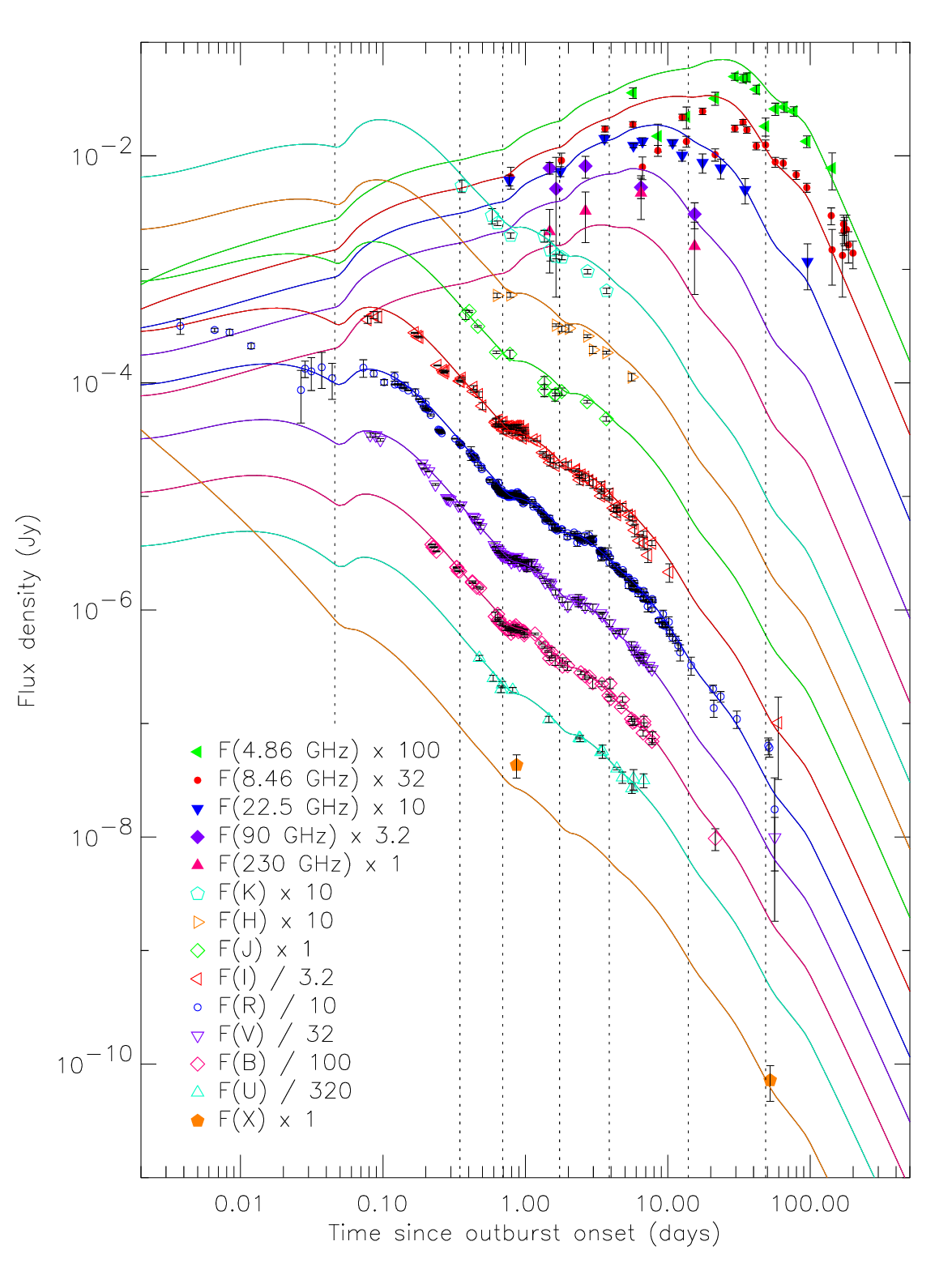}}
\caption{\label{LCmodel}
Multiband light curves from radio to X-rays (for the time interval
0.01 - 250 days after the burst onset) fitted with the multiple energy
injection model. Seven energy injection episodes (see Table~\ref{tabmodel}),
indicated by vertical lines, can  account for the observed behaviour. 
The visible and NIR observations are corrected for extinction
and the \textit{U}-band also for Lyman-$\alpha$ blanketing.
\label{figmodel}}
\end{center}
\end{figure*}

\begin{figure}[h]
\begin{center}
  {\includegraphics[width=\hsize]{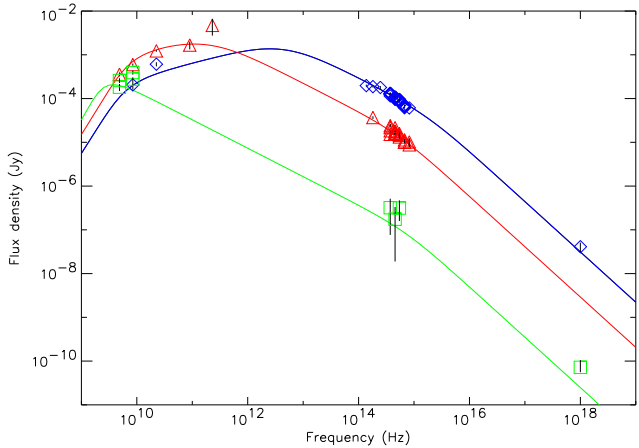}}
\caption{\label{SEDmodel}
Evolution of the GRB\,021004 afterglow SFDs at 0.8 (diamonds), 
6 (triangles) and 60 days (squares) after the burst.\label{seds}}
\end{center}
\end{figure}

Fig.~\ref{figmodel} shows  all the observational data along  with the light
curves predicted  by the  model for each  band.  Fig.~\ref{seds}  shows the
evolution of the afterglow multiband SFD at three epochs. As predicted by 
the model, we observe an evolution of the peak frequency from infrared to
radio as the afterglow decays. We note the excellent \textit{U}-band light 
curve prediction (not used for the fit) once the Lyman-$\alpha$ blanketing
is introduced.

\begin{table}[h]
\begin{center}
\caption{The 7-injection episodes model parameters. The injection 
energies $E_1$ to $E_7$ are in units of $E_0$, the initial energy. 
Other model parameters are: the initial Lorentz factor $\Gamma_{0}$, the 
ambient density $n_{0}$, the half opening angle $\theta_{0}$, the line 
of sight angle $\theta_{\nu}$, the electron energy index $p$, the fraction 
of internal energy stored in electrons after acceleration $\epsilon_{e}$ 
and the fraction of internal energy stored in the form of magnetic field 
$\epsilon_{B}$.
\label{tabmodel}}
\scriptsize{
\begin{tabular}{l l}
\noalign{\smallskip} 
\hline
\hline 
\noalign{\smallskip}Parameter & Value \\
\noalign{\smallskip} 
\hline \noalign{\smallskip}
$E_{0}$ (10$^{50}$erg)  & 1.5  \\
$E_{1}$ (0.046 days)   & 2.2  \\
$E_{2}$ (0.347 days)   & 0.7  \\
$E_{3}$ (0.694 days)   & 4.6  \\
$E_{4}$ (1.736 days)   & 10.0 \\
$E_{5}$ (3.877 days)   & 8.6  \\
$E_{6}$ (13.89 days)   & 10.0 \\
$E_{7}$ (48.61 days)   & 15.0 \\ 
$\Gamma_{0}$          & 770  \\
$n_{0} (cm^{-3})$     & 60.0 \\
$\theta_{0}$          & 1$\fdg$8 \\
$\theta_{\nu}$        & 0.8$\theta_{0}$ \\
$p$                   & 2.2  \\
$\epsilon_{e}$         & 0.35 \\
$\epsilon_{B}$         & 1.7$\times$10$^{-4}$\\
\noalign{
\smallskip} \hline \hline\end{tabular}
} \normalsize \rm
\end{center}
\end{table}

 \subsection{The host galaxy}
\label{host} 

In order to study the SFDs and constrain the model of the afterglow
we need  to isolate  the flux produced  by the  afterglow from that  of the
underlying  host galaxy.   For the  study  of the  host galaxy  we use  the
$BVIJ$-band magnitudes measured when  the contribution of the afterglow was
negligible, between $\sim62$(\textit{B}) and $\sim454$(\textit{J}) days 
after the burst.

\begin{figure}[ht!]
\centering\includegraphics[width=\hsize]{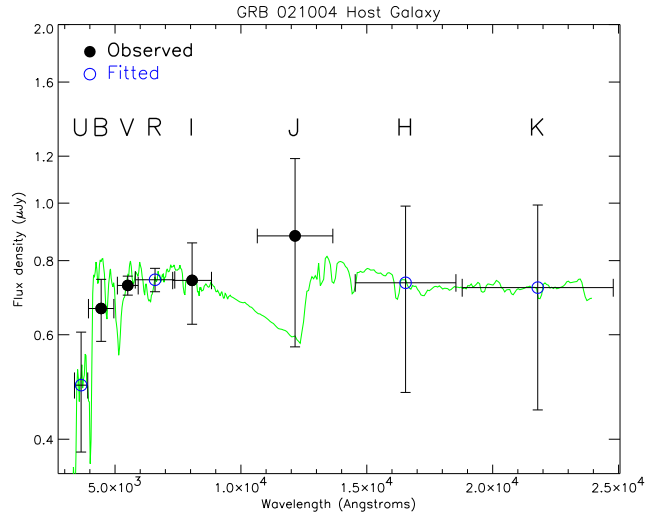}
\protect\caption[ ]{A fit to the photometric points of the GRB\,021004 
host galaxy yielding a starburst galaxy template. The \textit{H}-band 
limit of 21.5 magnitude ($\sim 2.7 \mu$ Jy) is not plotted. The filled 
points are our observations while the open ones are obtained from the 
fit of the galaxy. The flux density is represented in logarithmic scale.
\label{host021004}}
\end{figure}

The  fit of  the host  galaxy SFD  is based  on HyperZ  (Bolzonella  et al.
\cite{Bolz00}). The  fitting assumes Solar  metallicity, a Miller  \& Scalo
(\cite{Mill79})  initial mass  function  (IMF), and  an  SMC extinction  law
(Pr\'evot  et al.  \cite{Prev84}).   The  best fit  ($\chi^{2}/d.o.f.$  = 0.1)  is
obtained with a $\sim15$ Myr starburst galaxy with an absolute magnitude of
$M_{B} = -22.0\pm0.3$ and an  intrinsic extinction of $A_{V} = 0.06\pm0.08$
(see Fig.~ \ref{host021004}).

For  the subtraction of  the host  galaxy colours  in all  the optical/NIR
bands, it is  necessary to predict its magnitudes  in the $URHK$-bands, for
which no  photometric information is  available. Convolving the  spectra of
the fitted galaxy with standard  optical and infrared filters, a prediction
of  those  magnitudes  is  possible (see Table~\ref{hostmag}), assuming  the 
transformations given  by  Fukugita et  al. (\cite{Fuku95}). The 
\textit{U}-band must be corrected for Lyman-$\alpha$ blanketing as described 
in Sect. 4.3. In order to calculate the errors of the estimated magnitudes, 
a Monte Carlo method is used, in which the fitting of the galaxy is repeated 
with randomly modified input $BVIJ$  magnitudes 
(Gaussianly weighted) in the measured error range.

\begin{table}[h]
\begin{center}
\caption{Magnitudes  for the GRB\,021004 host galaxy. The  values marked
with $^{\star}$ are measured values,  while the rest are predictions obtained 
from the fitted template. The magnitudes are  given in the Vega system and 
are not corrected for Galactic reddening. \label{hostmag}}  \scriptsize{
\begin{tabular}{c l}\noalign{\smallskip} 
\hline\hline 
\noalign{\smallskip}
Band&Magnitude\\
\noalign{\smallskip} 
\hline \noalign{\smallskip}
U & 24.49 $\pm$ 0.28\\
B & 24.65 $\pm$ 0.13$^{\star}$\\
V & 24.45 $\pm$ 0.04$^{\star}$\\
R & 24.21 $\pm$ 0.05\\
I & 23.82 $\pm$ 0.17$^{\star}$\\
J & 23.15 $\pm$ 0.38$^{\star}$\\
H & 22.92 $\pm$ 0.31\\
K & 22.42 $\pm$ 0.37\\
\hline
\multicolumn{2}{l}{$\star$ Measured values. }\\
\noalign{
\smallskip} \hline \hline\end{tabular}
} \normalsize \rm
\end{center}
\end{table}

\section{Discussion}
\label{discussion}

We show that the multiwavelength observations can be satisfactorily 
reproduced in the context of the refreshed shock model. It is capable  
of reproducing the "bumpy"  behaviour of the entire light curve of 
this event, not  only the visible bands, but in a spectral range that 
spans from radio to X-rays. In addition, the  variations in the 
polarization of the  afterglow are naturally explained in  the 
framework of the energy injections (Bj\"ornsson et al. \cite{Bjor04}).
The number of required model parameters can be quite large, and depends on 
the structure of the light curve and modelling detail required. 

The bumpy light curve behaviour may also be explained by the patchy
shell model (Nakar et al.  \cite{Naka03}), or by density variations in
the surrounding medium (Lazzati et al.  \cite{Lazz02}), although in
the latter case, simultaneously accounting for the polarization
measurements appears to be problematic.  As in the refreshed shock
model, the number of required parameters in these models also
increases with the amount of structure in the light curve.

The  afterglow  model  assumes  an  adiabatic expansion,  so  the  proposed
scenario might  not be valid at  very early times, when  this assumption 
does not  apply.  Additionally, at early  time, there can also  be a strong
contribution from the early  reverse shock to  the light curve.   This might
explain the excess of \textit{R}-band flux  observed during the first 
minutes that followed the burst. The very last points of our dataset for 
GRB\,~021004 may indicate a transition to a non-relativistic expansion regime. 
We have not included all of the relevant modifications required to capture 
such a transition in detail and our model results may therefore be inaccurate 
at very late times.

Although  the X-ray  observations are  very limited  (only  two measurements),
there seems to be an excess in the observed flux as compared to the model. 
This could be  due to inverse Compton effect as seen in other GRBs (Harrison 
et al. \cite{Harr01}, in\' \rm t Zand  et al. \cite{Zand01}, Castro-Tirado et 
al. \cite{ajct03}), an effect not considered in our modelling. The correction 
for the Lyman-$\alpha$ blanketing in the \textit{U}-band that we introduced in Sect. 
4.3 shifts the photometric points consistently with the prediction of the GRB 
model.

The inferred host galaxy  extinction ($A_{V}$), dominant stellar age ($\sim
15$  Myr)  and galaxy  type  (starburst)  are  consistent with the  findings
reported by  Fynbo et al.  (\cite{Fynb05})  for GRB\,021004.   The age and
the extinction are  also consistent with the ones derived  for GRB hosts in
general, being  similar to young  starburst galaxies present in  the Hubble
Deep  Field sample (Christensen et al. \cite{Chri04}).   
However,   the  $B$-band  absolute  magnitude   of the host galaxy of  
GRB\,021004 ($M_{B}\simeq-22.0$) is  brighter than  the 10 hosts  present 
in  the above mentioned sample.

\section{Conclusions}

   Due to the early detection and rapid follow-up of GRB\,021004 we have 
had the opportunity of obtaining a very complete dataset concerning temporal 
range, wavelength coverage and sample  density.  This has allowed us to 
introduce important constrains on the models capable to explain the 
bumps present in the afterglow light curve.

In our analysis we assume several energy injection episodes to explain
the light curve. A reasonable scenario includes an initial burst
followed by 7 refreshed shocks. These add up to a total burst energy
of 7.8x10$^{51}$ ergs, that were emitted through a collimated jet with
an initial half-opening angle of 1$\fdg$8, pointing almost
directly towards us.

  A study of the photometric data of the host galaxy of GRB\,021004 reveals a
bright ($M_{B} = -22.0\pm0.3$) starburst galaxy with low extinction ($A_{V}
= 0.06\pm0.08$).

   Further tests of afterglow  models with this multiwavelength dataset are
encouraged.    Future   efforts   should   be   aimed   towards   obtaining
multiwavelength  photometry and  polarimetric observations  in order  to be
able to discriminate between the different models.

\section*{Acknowledgements}
We acknowledge the generous allocation of observing time by different Time
Allocation Committees at several observatories spread world-wide.
Partly based on observations carried out with the IRAM Plateau
de Bure Interferometer. IRAM is supported by INSU/CNRS (France), MPG
(Germany) and ING (Spain).
This research has been partially supported by the Spain's Ministerio de 
Ciencia y Tecnolog\' ia under programmes ESP2002-04124-C03-01 and AYA2004-01515 
(including FEDER funds). A. de Ugarte Postigo acknowledges support from a
FPU grant from Spain's Ministerio de Educaci\'on y Ciencia. G. J\'ohannesson, 
G. Bj\"ornsson and E.H. Gudmundsson acknowledge support from a special grant 
from the Icelandic Research Council. V.V. Sokolov and T. A. Fatkhullin were 
supported by Russian Foundation for Basic Research, grant No 01-02-17106.
We acknowledge our anonymous referee for constructive comments.

\begin{longtable}{cccccc}
\caption{Optical and NIR observations carried out for the GRB\,021004
afterglow. The magnitudes are in the Vega system and not corrected
for Galactic reddening. \label{visnir}}\\

\hline\hline
    Date UT      & Telescope &  Filter  & Texp (s)  &  Mag  & ErMag\\
\hline
\endfirsthead
\caption{continued.}\\
\hline\hline
    Date UT      & Telescope &  Filter  & Texp (s)  &  Mag  & ErMag\\
\hline
\endhead
\hline
\endfoot
2002 Oct 4.9792  & 2.2CAHA   & U       &  900   & 19.15 & 0.07 \\
2002 Oct 5.0986  & 2.2CAHA   & U       &  900   & 19.59 & 0.08 \\
2002 Oct 5.1785  & 2.2CAHA   & U       &  900   & 19.83 & 0.09 \\
2002 Oct 5.9653  & 2.2CAHA   & U       &  1200  & 20.48 & 0.08 \\
2002 Oct 6.9284  & 2.2CAHA   & U       &  1800  & 20.89 & 0.08 \\
2002 Oct 7.9767  & 2.2CAHA   & U       &  600   & 21.16 & 0.17 \\
2002 Oct 9.2990  & 1.0USNO   & U  &5$\times$1800& 21.70 & 0.14 \\
2002 Oct 10.244  & 1.0USNO   & U       &  1800  & 21.73 & 0.25 \\
2002 Oct 11.269  & 1.0USNO   & U  &4$\times$1800& 21.76 & 0.17 \\

\hline
2002 Oct 5.4890  & 1.0USNO   & B       &  1200  & 20.30 & 0.05 \\
2002 Oct 5.9844  & 2.2CAHA   & B       &  600   & 20.86 & 0.05 \\
2002 Oct 6.0300  & 1.52Loiano& B       &  1800  & 20.73 & 0.11 \\
2002 Oct 6.9537  & 2.2CAHA   & B       &  1800  & 21.17 & 0.04 \\
2002 Oct 6.9965  & 1.52Loiano& B  &2$\times$2400& 21.42 & 0.09 \\
2002 Oct 7.9522  & 2.2CAHA   & B       &  1200  & 21.41 & 0.06 \\
2002 Oct 9.3160  & 1.0USNO   & B   &5$\times$900& 21.74 & 0.06 \\
2002 Oct 10.115  & 3.5TNG    & B   &2$\times$300& 22.12 & 0.05 \\
2002 Oct 11.287  & 1.0USNO   & B  &4$\times$1200& 22.20 & 0.11 \\
2002 Oct 26.050  & 4.2WHT    & B  &7$\times$300 & 24.28 & 0.13 \\
2002 Nov 27.003  & 2.5INT    & B  &10$\times$600& 24.54 & 0.05 \\
2002 Dec  5.762  & 6.0SAO    & B       &   3600 & 24.65 & 0.13 \\

\hline
2002 Oct 4.5857  & 0.6MOA    & Blue(V) &  180   & 16.90 & 0.05 \\
2002 Oct 4.5934  & 0.6MOA    & Blue(V) &  180   & 16.91 & 0.04 \\
2002 Oct 4.6002  & 0.6MOA    & Blue(V) &  180   & 17.01 & 0.03 \\
2002 Oct 4.9896  & 2.2CAHA   & V       &  300   & 18.93 & 0.05 \\
2002 Oct 5.1090  & 2.2CAHA   & V       &  300   & 19.32 & 0.05 \\
2002 Oct 5.1882  & 2.2CAHA   & V       &  300   & 19.54 & 0.05 \\
2002 Oct 5.4780  & 1.0USNO   & V       &  600   & 19.71 & 0.04 \\
2002 Oct 5.5232  & 0.6MOA    & Blue(V) &2$\times$600& 19.77 & 0.09 \\
2002 Oct 5.5330  & 0.6MOA    & Blue(V) &  600   & 19.76 & 0.13 \\
2002 Oct 5.5593  & 0.6MOA    & Blue(V) &2$\times$600& 19.71 & 0.07 \\
2002 Oct 5.6060  & 0.6MOA    & Blue(V) &3$\times$600& 19.74 & 0.07 \\
2002 Oct 5.9920  & 2.2CAHA   & V       &  300   & 19.74 & 0.07 \\
2002 Oct 6.0110  & 1.52Loiano& V       &  1200  & 20.18 & 0.06 \\
2002 Oct 6.1240  & 1.52Loiano& V       &  900   & 20.15 & 0.31 \\
2002 Oct 6.8650  & 1.52Loiano& V       &  1800  & 20.52 & 0.13 \\
2002 Oct 6.9410  & 1.52Loiano& V  &2$\times$1800& 20.75 & 0.08 \\
2002 Oct 6.9595  & 2.2CAHA   & V       &  600   & 20.55 & 0.05 \\
2002 Oct 7.9668  & 2.2CAHA   & V       &  600   & 20.85 & 0.06 \\
2002 Oct 9.3250  & 1.0USNO   & V   &5$\times$600& 21.20 & 0.05 \\
2002 Oct 10.348  & 1.0USNO   & V   &5$\times$600& 21.60 & 0.08 \\
2002 Oct 11.298  & 1.0USNO   & V   &4$\times$600& 21.75 & 0.10 \\
2002 Nov 29.811  & 6.0SAO    & V       &   2250 & 24.43 & 0.17 \\
2002 Dec  5.698  & 6.0SAO    & V       &   3600 & 24.13 & 0.09 \\
2003 Sep  17.073 & 4.2WHT    & V   &5$\times$900& 24.45 & 0.04 \\

\hline
2002 Oct 4.9965  & 2.2CAHA   & Rc      &  300   & 18.55 & 0.03 \\
2002 Oct 5.1146  & 2.2CAHA   & Rc      &  300   & 18.96 & 0.02 \\
2002 Oct 5.1938  & 2.2CAHA   & Rc      &  300   & 19.12 & 0.03 \\
2002 Oct 5.4700  & 1.0USNO   & Rc      &  600   & 19.35 & 0.04 \\
2002 Oct 5.5000  & 1.0USNO   & Rc      &  600   & 19.31 & 0.04 \\
2002 Oct 5.9790  & 1.52Loiano& Rc      &  1200  & 19.67 & 0.06 \\
2002 Oct 5.9976  & 2.2CAHA   & Rc      &  300   & 19.73 & 0.03 \\
2002 Oct 6.0500  & 1.52Loiano& Rc      &  1200  & 19.87 & 0.06 \\
2002 Oct 6.0620  & 3.5TNG    & Rc  &2$\times$120& 19.72 & 0.04 \\
2002 Oct 6.8140  & 1.52Loiano& Rc &3$\times$1800& 20.20 & 0.08 \\
2002 Oct 6.9687  & 2.2CAHA   & Rc      &  600   & 20.13 & 0.03 \\
2002 Oct 7.9906  & 2.2CAHA   & Rc      &  600   & 20.39 & 0.06 \\
2002 Oct 9.3060  & 1.0USNO   & Rc  &4$\times$600& 20.87 & 0.06 \\
2002 Oct 10.105  & 3.5TNG    & Rc  &2$\times$180& 21.05 & 0.05 \\
2002 Oct 10.356  & 1.0USNO   & Rc  &5$\times$600& 21.14 & 0.07 \\
2002 Oct 11.305  & 1.0USNO   & Rc  &4$\times$600& 21.35 & 0.10 \\
2002 Nov 29.693  & 6.0SAO    & Rc      &  2700  & 24.29 & 0.18 \\

\hline
2002 Oct 4.5823  & 0.6MOA    & Red(Ic) &  180   & 16.19 & 0.08 \\
2002 Oct 4.5900  & 0.6MOA    & Red(Ic) &  180   & 16.09 & 0.08 \\
2002 Oct 4.5968  & 0.6MOA    & Red(Ic) &  180   & 16.13 & 0.12 \\
2002 Oct 5.0035  & 2.2CAHA   & Ic      &  600   & 18.09 & 0.08 \\
2002 Oct 5.1215  & 2.2CAHA   & Ic      &  600   & 18.45 & 0.08 \\
2002 Oct 5.2017  & 2.2CAHA   & Ic      &  600   & 18.62 & 0.08 \\
2002 Oct 5.2750  & 1.55USNO  & Ic      &  900   & 18.68 & 0.02 \\
2002 Oct 5.3970  & 1.55USNO  & Ic      &  900   & 18.71 & 0.01 \\
2002 Oct 5.4620  & 1.0USNO   & Ic      &  600   & 18.79 & 0.06 \\
2002 Oct 5.5149  & 0.6MOA    & Red(Ic) &2$\times$600& 18.90 & 0.13 \\
2002 Oct 5.5247  & 0.6MOA    & Red(Ic) &  600   & 18.85 & 0.17 \\
2002 Oct 5.5510  & 0.6MOA    & Red(Ic) &2$\times$600& 18.61 & 0.11 \\
2002 Oct 5.6067  & 0.6MOA    & Red(Ic) &3$\times$600& 18.98 & 0.13 \\
2002 Oct 5.9940  & 1.52Loiano& Ic      &  900   & 19.32 & 0.08 \\
2002 Oct 6.0069  & 2.2CAHA   & Ic      &  900   & 19.24 & 0.08 \\
2002 Oct 6.0580  & 3.5TNG    & Ic      &2$\times$120& 19.22 & 0.04 \\
2002 Oct 6.9120  & 1.52Loiano& Ic      &  1200  & 19.70 & 0.10 \\
2002 Oct 6.9811  & 2.2CAHA   & Ic      &  900   & 19.60 & 0.07 \\
2002 Oct 8.3310  & 1.55USNO  & Ic  &2$\times$900& 20.09 & 0.02 \\
2002 Oct 9.3120  & 1.0USNO   & Ic  &4$\times$480& 20.29 & 0.08 \\
2002 Oct 10.363  & 1.0USNO   & Ic  &5$\times$600& 20.78 & 0.13 \\
2002 Oct 11.313  & 1.0USNO   & Ic  &4$\times$600& 21.05 & 0.13 \\
2002 Dec  3.000  & 6.0SAO    & Ic      &  2640  & 23.77 & 0.19 \\
2003 Dec 28.888  & 2.5NOT    & Ic &14$\times$300& 23.82 & 0.17 \\

\hline
2002 Oct 4.8847  & 1.5Tirgo  & J       &  1680  & 16.74 & 0.10 \\
2002 Oct 5.1167  & 1.5Tirgo  & J       &  1920  & 17.90 & 0.23 \\
2002 Oct 5.8514  & 1.5Tirgo  & J       &  1680  & 18.06 & 0.39 \\
2004 Jan  5.805  & 3.5CAHA   & J       &  7260  & 23.15 & 0.38 \\

\hline
2004 Jan 7.2775  & 3.5CAHA   & H       &  6120  &$>21.5$& ---  \\ 

\hline
2002 Oct 4.8622  & 1.5Tirgo  & Ks      &  1740  & 15.30 & 0.14 \\
2002 Oct 5.0882  & 1.5Tirgo  & Ks      &  2040  & 15.95 & 0.17 \\
2002 Oct 5.8743  & 1.5Tirgo  & Ks      &  1920  & 16.12 & 0.24 \\
2002 Oct 6.0014  & 1.5Tirgo  & Ks      &  3480  & 16.71 & 0.21 \\

\end{longtable}
\end{document}